\documentclass[sensors,article,submit,moreauthors,pdftex]{Definitions/mdpi}

\firstpage{1}
\pubvolume{xx}
\issuenum{1}
\articlenumber{5}
\pubyear{2019}
\copyrightyear{2019}
\history{Received: date; Accepted: date; Published: date}
\renewcommand{\linenumbers}{}

\usepackage{lscape}

\Title{Communication Technologies for Smart Grid: A Comprehensive Survey}

\Author{Fredrik Ege Abrahamsen  $^{1,*}$, Yun Ai $^{2}$ and Michael Cheffena  $^{2}$}

\AuthorNames{Fredrik Ege Abrahamsen, Yun Ai and Michael Cheffena}

\address{%
$^{1}$ \quad Faculty of Information Technology and Electrical Engineering, Norwegian University of Science and Technology, 2815 Gjøvik, Norway.\\
$^{2}$ \quad Faculty of Engineering, Norwegian University of Science and Technology, 2815 Gjøvik, Norway.}

\corres{Correspondence: fredrik.e.abrahamsen@ntnu.no}

\abstract{With the ongoing trends in the energy sector such as vehicular electrification and renewable energy, smart grid is clearly playing a more and more important role in the electric power system industry. One essential feature of the smart grid is the information flow over the high-speed, reliable and secure data communication network in order to manage the complex power systems effectively and intelligently. Smart grids utilize bidirectional communication to function where traditional power grids mainly only use one-way communication. The communication requirements and suitable technique differ depending on the specific environment and scenario. In this paper, we provide a comprehensive and up-to-date survey on the communication technologies used in the smart grid, including the communication requirements, physical layer technologies, network architectures, and research challenges. This survey aims to help the readers identify the potential research problems in the continued research on the topic of smart grid communications.}

\keyword{review, survey, smart grid, smart grid technologies, smart grid communication, wireless communications, wired communication, smart grid security}

\begin{document}
%%%%%%%%%%%%%%%%%%%%%%%%%%%%%%%%%%%%%%%%%%

%%%%%%%%%%%%%%%%%%%%%%%%%%%%%%%%%%%%%%%%%%
\section{Introduction}
Today’s method for generation and distribution of electric power was designed and constructed in the last century and has largely remained unchanged since then. The traditional power grids are primarily radial, and built for centralized power generation. Reliability is ensured by having excessive capacity and one-way power flow from the power plant to the consumer through high voltage transmission lines, often over long distances. With the demand for electric energy continuously increasing, and the existing conventional grid being at the end of its life cycle. Increasing amount of distributed renewable energy sources (RES) and energy storage systems (ESS) require new ways of managing and controlling the power grid and distributing the power in a more efficient, effective environmentally sustainable and economical manner. The next generation power grid is often referred to as smart grids (SGs). Smart grids are achieved by overlaying a hierarchical communication infrastructure on the power grid infrastructure \cite{Zhou_summary,priya_performance_2014,farhangi2009path}.  \\
Since January 1 2019, most end-users in Norway should have installed smart electricity meters to manage automated meter readings (AMR). In January 2019, 97 \% or 2.9 million meters were installed in Norway, among which 2.5 million of these are private households and holiday homes \cite{nve_smartstrommalere,nve2019smarte}. In the European Union (EU), it was committed by the member states to achieve a roll out of close to 200 million smart meters for electricity by 2020. About 71 \% of European consumers then will have a smart electricity meter installed \cite{RN29}.  Globally, it is expected that 800 million smart meters will be installed by 2020 \cite{RN46}. The installation of these metering devices can be seen as one of the first steps toward a smarter grid system, as implementing a smart grid is not a one-time event rather than an evolutionary process. The smart meters have the ability to collect and report consumption data to the utilities provider several times per hour, rather than the consumer having to report every month manually. The smart metes also open up for the consumer to feed the grid with electricity from, i.e. solar panels or electric vehicles. Other possibilities with the smart metering are a higher degree of monitoring and control of the grid, automatic fault detection and reports \cite{iea_tech_roadmap, Zhou_summary}.\\

\begin{figure}[H]
    \centering
    \includegraphics[width=\textwidth]{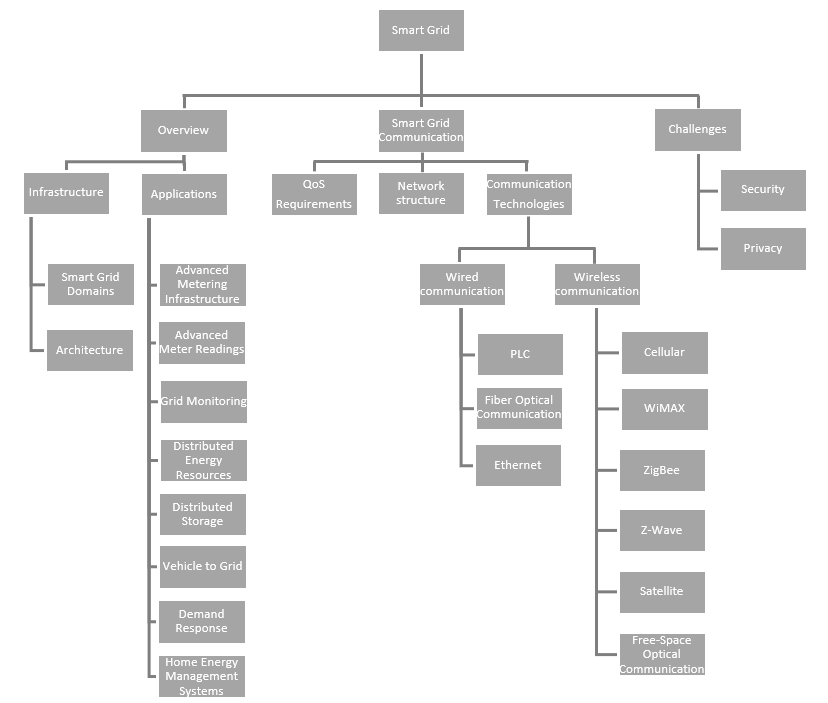}
    \caption{The structure of the paper}
    \label{fig:structure}
\end{figure}

This remainder of the paper is organized as follows: Section 2 gives an overview of smart grid infrastructure, domains, architecture and applications. Section 3 presents smart grid communication technologies and network structures. Section 4 addresses challenges of smart grid communications, and privacy and security of smart grid communication. The organization of this paper is summarized in Figure \ref{fig:structure}.

%\textbf{\color{red}Motivation for the paper, the contribution etc.}

\section{Overview of Smart Grid}
Communication plays an important role in SG, as one of the most significant differences between traditional grids and SG are the two-way communication and the potentials this enables (i.e., distributed smart sensors, distributed power generation, real time measurements and metering infrastructure, monitoring systems, and fast response require reliable communication and information exchange). Information exchanges is of great importance for the SG to provide reliable power generation and distribution. Following is an overview of smart grid infrastructure, domains, network architecture and smart grid applications.

%\textbf{\color{red}Incorporate the role of information exchange in the applications}

\subsection{Smart Grid Infrastructure}
Both international and national organizations have developed roadmaps, defined standards and definitions on what makes a power grid a smart grid \cite{iea_tech_roadmap,iec_roadmap,nist_roadmap_1,nist_roadmap_3,dke_roadmap,sg_austria_roadmap,itu_sg_repository}.
There are not one single definition on what a smart grid is. However, it is common in the SG definitions to emphasize on the communication for measurements, monitoring, management and control. The communication plays an essential role in providing reliable, efficient and secure power generation, transmission and distribution. The communication systems provide information exchange between the distributed sensing equipment,  monitoring systems, and data management systems. These solutions all require fast communications as the generation, delivery and consumption all happening at the same time. With the introduction of distributed energy sources and energy storage systems, the importance of fast and reliable communication increases. The expectations from end-users also change, with real-time information on electricity prices, customers feeding the grid with electricity and electric vehicles acting as batteries in the grid. A key goal for smart grids are reduced cost and environmental impact, and maximizing reliability, resilience and stability \cite{iea_tech_roadmap}.
The smart meter is a key component of the smart grid infrastructure, and part of the advanced metering infrastructure (AMI). AMI is responsible for enabling a reliable and secure high speed two way communication between smart meters at the end-user, and data control centers at the utilities companies for monitoring and control \cite{chren2016smart,iec_roadmap, gozde20154g}.
The full benefit of the smart grid infrastructure is achieved when smart meters, sensors and measuring devices located throughout the power grid communicate in order to ensure stability, detect, predict and prevent faults, forecast load changes and facilitate demand response \cite{goel2015smart}.
Table \ref{tab:comparison_grids} shows the main differences between traditional grid and a Smart grid.

\begin{table}[H]
    \centering
    \caption{Comparison of traditional power grid and smart power grid \cite{farhangi2009path}.}
    \begin{tabular}{l|l|l}
        \toprule
                            & \textbf{Traditional grid} & \textbf{Smart grid}  \\
        \midrule
        Information flow    & One-way communication & Two-way communication \\
        Power generation    & Centralized power generation & Distributed power generation \\
        Grid topology       & Radial    & Network \\
        Integration of distributed & Low degree & High degree \\
        energy sources  \\
        Sensors             & Low degree & High degree \\
        Monitoring          & Manual monitoring & Self-monitoring \\
        Outage recovery     & Manual restoration & Self-reconfiguration \\
        Testing             & Manual & Remote \\
        Ability to control & Limited & Pervasive \\
        Efficiency          & Low   & High \\
        Environmental impact & High & Low \\
        \bottomrule
    \end{tabular}
    \label{tab:comparison_grids}
\end{table}

\subsubsection{Smart Grid Domains}
\begin{figure}[H]
    \centering
    \includegraphics[width=\textwidth]{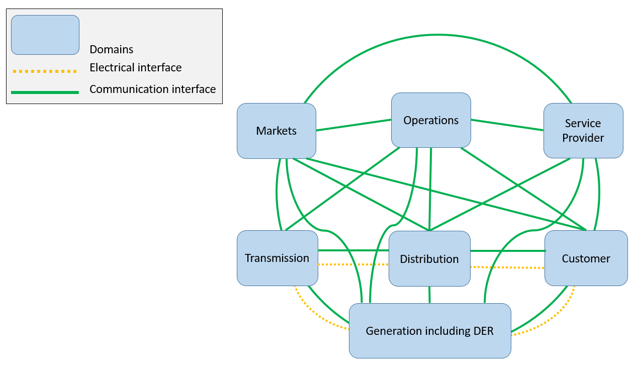}
    \caption{Domains in smart grid as proposed in NIST Smart Grid Framework 4.0 \cite{nist_roadmap_4}}
    \label{fig:sg_domains}
\end{figure}

Smart grids are large systems, which include different stakeholders and domains \cite{kuzlu2014communication}. To structure the different areas of a smart grid environment, The National Institute of Standards and Technology (NIST) \cite{nist_roadmap_1} proposed seven domains of smart grid with electrical interfaces and communication interfaces in its conceptual model for smart grid information networks in 2009. The conceptual model has later been updated with more communications and electrical interfaces to better reflect the increase in distributed energy sources and automation of distribution systems \cite{nist_concept_model,nist_roadmap_4}. Table \ref{tab:comparison_tg_sg} shows the definition of these domains and Figure \ref{fig:sg_domains} displays the domains and interfaces. The domains are: customer, distribution, transmission, generation including distributed energy resources (DER), markets, operations and service providers. The first four being related to transmission of electricity on the power grid. Table \ref{tab:comparison_tg_sg} gives the descriptions of the NIST model.

\begin{table}[H]
\caption{Smart Grid domains, electrical and communication interface  \cite{hossain2012smart}.}
\label{tab:comparison_tg_sg}
\centering
\begin{tabular}{lll}
\toprule
\textbf{Domain} & \textbf{Communication interface} & \textbf{Electrical interface} \\
\midrule
Market                           & \begin{tabular}[c]{@{}l@{}}Service provider, Operations, Generation,\\ Transmission, Distribution, Customer\end{tabular} & None  \\
\midrule
Operations                       & \begin{tabular}[c]{@{}l@{}}Markets, Service provider, Transmission, \\ Distribution, Customer, Generation\end{tabular}   & None \\
\midrule
Service provider                 & \begin{tabular}[c]{@{}l@{}}Markets, Operations, Customer, \\ Distribution, Generation\end{tabular}& None\\
\midrule
Transmission& \begin{tabular}[c]{@{}l@{}}Markets, Operations  Generation, \\ Distribution\end{tabular} & Generation, Distribution \\
\midrule
Distribution & \begin{tabular}[c]{@{}l@{}}Operations, Transmission, Customer, \\ Service Provider\end{tabular} & Transmission, Customer \\
\midrule
Customer                         & \begin{tabular}[c]{@{}l@{}}Markets, Operations, Service provider,\\ Distribution\end{tabular}& Distribution, Generation\\
\midrule
Generation incl. DER             & \begin{tabular}[c]{@{}l@{}}Markets, Operations, Transmission, \\ Customer\end{tabular}                                   & Transmission, Customer                         \\
\bottomrule
\end{tabular}
\end{table}

Each component of Figure \ref{fig:sg_domains} is described below.
\begin{itemize}
    \item \textbf{Market domain:} Grid assets and services are bought and sold within the domain. The market domain handles actors such as market management, wholesale, trading and retailing. The market domain communicates with all other domains in the smart grid. Communication between market domain and the energy supplying domains are critical, due to the need for efficient matching of production and consumption \cite{nist_roadmap_4}.
    \item \textbf{Operations domain:} The domain is responsible for operations of the grid. Including monitoring, control, fault detection and management, grid maintenance and customer support. These are typically the responsibilities of the utilities today. With smart grid more of these responsibilities will move over to service providers \cite{nist_roadmap_1,nist_roadmap_3}.
    \item \textbf{Service provider domain:} Actors in the domain support business processes of power producers, distributors and customers. Ranging from utility services such as billing to management of energy use and generation. Communication interface is shared with the generation, distribution, markets, operations and customer. The communication with operations domain is critical to ensure system control and situational awareness \cite{nist_roadmap_1,nist_roadmap_3}.
    \item \textbf{Generation domain:} The power generation domain is responsible for power generation in bulk or non-bulk quantities. This can be from for example fossil fuels, water, wind or solar. For the case of Norway, this is typically hydropower, these are grid-connected power generation stations. Power generation include distributed energy resources. Smart grids allow for end-users to also operate as producer of electrical energy, for premise use, storage or for resale. \cite{nist_roadmap_3,ma2013smart}. With smart grids, power generation is no longer limited to large fossil or hydroelectric power facilities feeding the transmission grid. Smart grids allow for smaller scale distribution-grid-connected power generation. This can be wind power parks, solar parks, photovoltaic panels mounted on end-users roof-tops, or electric vehicles feeding the grid \cite{nist_roadmap_3,samad2017controls}. Communication with the transmission and distribution domains are important to maintain energy delivery to customers \cite{nist_roadmap_1,nist_roadmap_3}.
    \item \textbf{Transmission domain:} The power transmission domain is responsible for the transfer of power from the power generation source to the distribution system. The transmission domain typically consists of transmission lines, substations, energy storage systems and measurement and control systems. The transmission system is typically monitored and controlled through supervisory control and data acquisition (SCADA) system which communicates with field and control devices throughout the transmission grid \cite{nist_roadmap_1,nist_roadmap_3}.
    \item \textbf{Distribution incl. DER domain:} This domain is the connection between the transmission and the customer domain. The distribution domain may include DERs located at customer or at grid operator. In a smart grid environment, the distribution domain communicates with the market domain due to the market domains potential to affect local power consumption and generation \cite{nist_roadmap_1,nist_roadmap_3,nist_roadmap_4}.
    \item \textbf{Customer domain:} The customer or end-user could be private, commercial or industrial. In addition to consume the energy, the customer could also generate, and feed the grid with excess energy or store energy. In cases where the customer generate and deliver energy consumer is referred to as a prosumer \cite{souri2014smart,nist_roadmap_4}.
\end{itemize}

Reliable communication is required for information exchange between the different domains to ensure reliable operations of the power grid and its applications.
Similar to NIST in the US, in Europe, the smart grid coordination group defined its smart grid architecture model \cite{iec_roadmap,sg_austria_roadmap,cenelec}. There are similarities between the two models, the domains are the same. In addition to domains this model is also divided in layers and zones.

This is a three dimensional model consisting of five interoperability layers (business, function, information, communication and components). The two dimensions are divided in domains (generation, transmission, distribution, DER, and customer premises), and zones (process, field, station, operation, enterprise and market).

\begin{table}[b]
\caption{Overview of smart grid communication layers}
\label{tab:sg_com_overview}
\centering
\begin{tabular}{|l|c|c|c|l|}
\hline
\textbf{Application Layer} & \multicolumn{2}{c|}{\textbf{\begin{tabular}[c]{@{}c@{}}Power Transmission and \\ Distribution Applications\end{tabular}}} & \multicolumn{2}{c|}{\textbf{Customer Applications}} \\ \hline
\rowcolor[HTML]{EFEFEF}
Communication Layer & Wide Area Network & \begin{tabular}[c]{@{}c@{}}Neighborhood Area \\ Network / Field Area \\ Network\end{tabular} & \multicolumn{2}{c|}{\cellcolor[HTML]{EFEFEF}\begin{tabular}[c]{@{}c@{}}Premise Area Network \\ (Home Area Network,\\ Building Area Network,\\ Industrial Area Network)\end{tabular}} \\ \hline
Power Control Layer & \multicolumn{4}{c|}{Power monitoring, control and management systems} \\ \hline
Power System Layer  & \begin{tabular}[c]{@{}c@{}}Power Generation \\ and Transmission\end{tabular} & Power Distribution  & \multicolumn{2}{c|}{Customer}  \\ \hline
\end{tabular}
\end{table}

\subsubsection{Architecture}
What separates smart grids from traditional electrical grids are the interaction and communication between different the domains. The smart grid infrastructure can be structured by dividing it in four layers: the application layer, the communication layer, the power control layer and the power system layer.
On the customer side, the application layer enables various applications such as home automation and real-time pricing. On the grid side: automation of grid, and power distribution applications.
The communication layer is important in distinguishing smart grids from traditional power grids, and in enabling smart grid applications. It is divided into three categories classified by geographic area (wide area network (WAN), neighborhood area network (NAN)/field area network (FAN) and the premise area network (PAN)). Depending on the type of network, different communication technologies are used. The power control layer enables management, control and monitoring of the power grid, utilizing equipment and such as switches, sensors and metering devices.
Power system layer handles power generation, transmission/distribution and the customer premises.

\subsection{Smart Grid Applications}
SG applications for monitoring and grid management include advanced metering infrastructure (AMI), advanced metering readings, distributed automation (DA), distributed generation (DG), distributed Storage, home energy management systems (HEMS), demand response (DR), and supervisory control and data acquisition. All depending on reliable wired and wireless communication interfaces to operate in the smart grid infrastructure. \\

\begin{figure}[t]
    \centering
    \includegraphics[width=15cm]{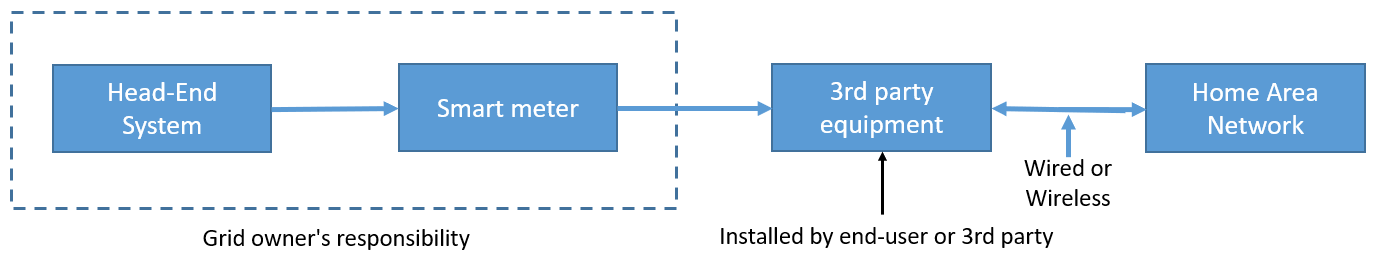}
    \caption{Smart metering architecture}
    \label{fig:sm_architecture}
\end{figure}

\noindent\textbf{Advanced metering infrastructure (AMI)}\\
SG is considered as one of the largest potential IoT network implementations with smart meters, wireless smart sensors placed throughout the grid and smart appliances communication with each other to ensure reliable and efficient power generation and distribution. The advanced metering infrastructure consists of physical and virtual components, including sensors, monitoring systems, smart meters, software, data management systems and communication networks.
AMI is responsible for collecting, analyzing and storing metering data sent from sensors and monitoring systems and smart meters at the end-user to the utility companies for billing, grid management and forecasting. SG interactions based on measured data and communication from sensor networks. \cite{kimani2019cyber,kabalci2019introduction}. \\

\noindent\textbf{Advanced meter readings}\\
The smart metering devices installed on the customer premises use different technologies for communicating. These vary depending on what manufacturer smart meter the utilities company are installing, and the application. Table \ref{tab:sm_reading_specs} lists the different parameters available from the end-users’ smart meters, and refresh rate in the Norwegian market \cite{nek_2018_han_personvern,nek_han_specification}.

\begin{table}[t]
    \caption{Norwegian smart meter HAN reading specification \cite{nek_han_specification}}
    \centering
    \begin{tabular}{llll}
        \toprule
        Refresh rate & Parameter & Unit & Data type \\
        \midrule
        2 sec.  & Active power + & kW & double-long-unsigned \\
        \midrule
        10 sec. & Meter ID && string \\
                & Meter type && string \\
                & Active power + & kW & double-long-unsigned \\
                & Active power - & kW & double-long-unsigned \\
                & Reactive power + & kVAr & double-long-unsigned \\
                & Reactive power - & kVAr & double-long-unsigned \\
                & Current x 3 & A & long-signed \\
                & Voltage x 3 & V & long-unsigned \\
        \midrule
        1 h     & Cumulative hourly active energy x 2 & kWh & double-long-unsigned \\
                & Cumulative hourly reactive energy x 2 & kVArh & double-long-unsigned \\
                & Time and date in meter && string \\
        \bottomrule
    \end{tabular}
    \label{tab:sm_reading_specs}
\end{table}

For large apartment buildings the metering devices can be connected to the master device by RS-485 \cite{bouwmeester2018towards}. As illustrated in Fig. \ref{fig:sm_architecture}, the metering devices can also be directly connected to the Head-End System (HES) through 3G/4G/5G or fiber network, so called end-to-end connection. The master device uses 3G/4G/5G, Ethernet, fiber optics or power line communication (PLC) to communicate with the head-end system at the utilities company. Inside the premise area, the smart meter communicates through the HAN-port, the communication is based on IEC 62056-7-8, with RJ45 connector and M-Bus interface. From this port, other third party equipment can be installed i.e. HEMS or household appliances \cite{han_grensesnitt}.\\

\noindent\textbf{Grid monitoring}\\
SCADA is responsible for monitoring of generation and power transmission, and the communication is performed on WANs. The SCADA functions are enhanced from traditional grid due to fast two-way communication and implementation of large numbers of sensors.
For transmission line monitoring, wireless smart sensor nodes are distributed along the transmission line, exchanging measurements to the neighboring nodes. The nodes forward the measurements to a central collection site over NAN or WAN. The central is connected to a base station with low latency, high bandwidth and low costs links \cite{faheem2018smart}.
To ensure uninterrupted power delivery continuous monitoring is required. Fast outage identification, management, and restoration systems  can be achieved by interfacing the outage management systems with SCADA, AMI and geographical information system. The combination of AMI and smart meters can give notifications or last gap reports to outage management systems before the customer notices the outage, thus helping in reducing trouble-shooting time and restoration time \cite{faheem2018smart}. These systems for status monitoring, of the smart grid infrastructure down to individual components helps to detect, predict and respond to faults faster. The result is better management, more accurate optimization of resources, better and faster identification of faults in the grid, reduction in troubleshooting-time, and improved reliability \cite{kimani2019cyber,kabalci2019introduction}.\\

\noindent\textbf{Distributed energy resources (DER)}\\
Distributed energy resources have a substantial potential at generating electricity at the load end. DER include solar photovoltaic panels, windpower and biomass. Two-way communication in the AMI enable the end user to sell surplus energy, and feeding it back to the power grid \cite{salinas2013dynamic, tushar2014three}.\\

\noindent\textbf{Distributed storage}\\
Distributed storage is an integral part of the smart grid infrastructure. Fast response to stability issues in the grid are dependent on fast and reliable communication links in the smart grid. Distributed storage in combination with DER can improve the utilization of renewable distributed energy resources \cite{logenthiran2011intelligent}.\\

\noindent\textbf{Vehicle to grid (V2G)}\\
It is clear that the electrification of vehicles is becoming an ongoing trend around the globe now, which implies the frequent interaction between power grid and vehicles in the future. Electric vehicles (EVs) with charges connected to a vehicle to grid network in the smart grid for exchanging power between the grid and the EV by utilizing the stored energy in the vehicle batteries to feed back to the grid when necessary. EVs in the power grid can be used for power balancing by providing fast response high power. EVs can reduce the energy demand in peak load hours by consuming, storing and returning energy when needed. EVs can also be used as back up power or in islanded operation if connection to the grid is not possible \cite{liu2013opportunities,bach2019parker}. These applications require bidirectional communication between the utilities and the EVs. \\

\noindent\textbf{Demand response (DR)}\\
AMI and communication between end-user and the utilities companies give ability for demand response from the consumer side, or utility side in predefined cases.
From the consumer side, demand response gives the end-user the ability to monitor its energy consumption and production where appropriate and cost at any given time, and for example alter habits to shift the demand to off-peak hours in response dynamic pricing programs such as time of use, real time pricing, critical peak timing or to incentive payment when grid reliability is low \cite{balakumar2017demand,mohagheghi2011impact,tsado2015resilient}. Demand response can also be an automated part of home energy management systems, where certain appliances or lighting can be turned off to reduce consumption  \cite{budka2016communication}. Demand side management or demand response can be used to reduce power constraint, shift peak load, reduce distribution losses and regulate voltage drops and avoid or postpone the need for building new power lines \cite{kimani2019cyber,balijepalli2011review}.\\

\noindent\textbf{Home energy management systems (HEMS)}\\
HEMS is used to enable demand response applications. HEMS systems permits the end-users to monitor, control, and manage the power consumption. These systems is comprised of smart appliances, sensors, smart meters and in-home displays, and include applications for for example home automation, temperature zone setting, water temperature, and controlling electricity use depending on real-time pricing information etc. Appliances and sensors connects to HEMS through sensor networks and to the utility companies AMI through the smart meter HAN interface \cite{han_grensesnitt,faheem2018smart}.

\section{Smart Grid Communication}
From the previous section we can see that SG is highly dependent on information flow and communication between different entities in different networks. Communication is one of enabling technologies of SG. As the number of sensors increases, the amount of data coming to the utility increases.

\subsection{Quality-of-service (QoS) Requirements for SG}
SG applications result in increased data, these applications have different QoS requirements. Secure bi-directional communication that satisfies the different SG application's QoS requirements are essential \cite{faheem2018smart}.
Control, management and automation applications such as demand response and substation automation require low latency and high reliability to ensure grid operation. Other applications such as meter readings can tolerate a higher latency, but still require high reliability \cite{faheem2018smart}.
With the different equipment interconnected in the SG, interoperability must be ensured for seamless communication \cite{nist_roadmap_3}. Interoperability must also be ensured for legacy and evolving communication protocols.

\begin{table}[H]
    \centering
    \begin{tabular}{llll}
    \toprule
         SG Application &  Data Rate & Latency & Reliability \\
    \midrule
         Smart Metering & Low & High & Medium \\
         SCADA & Medium & Low & Low \\
         Substation Automation & Low & Low & High \\
         DER & Medium & Low & High \\
         DR & Low & Low & High \\
    \bottomrule
    \end{tabular}
    \caption{Smart Grid QoS Requirements}
    \label{tab:QoS_Requirements}
\end{table}

%\textbf{\color{red}Latency, Bandwidth, Data rates, Throughput, Reliability, Data accuracy, Data validity, Accessibility, Interoperability.. ?} \\

\subsection{Communication Network Structure}
\begin{figure}[H]
    \centering
    \includegraphics[width=15cm]{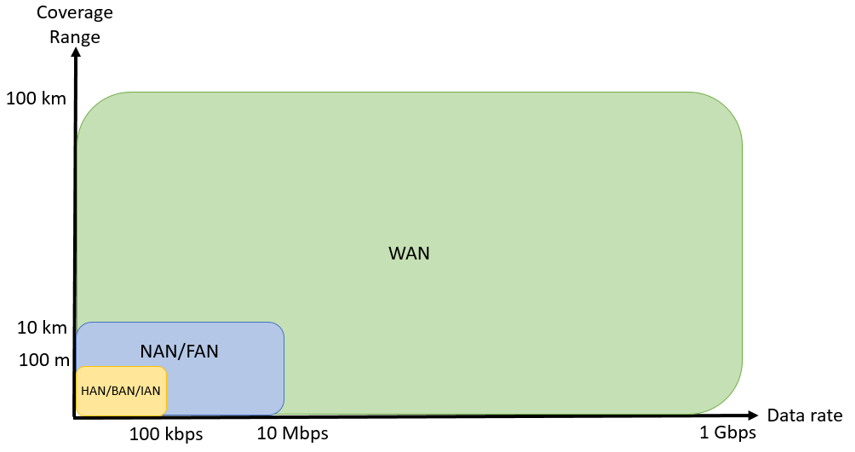}
    \caption{Data rate and communication range requirements in SG hierarchy}
    \label{fig:sg_range}
\end{figure}

A defined communications framework is necessary in this infrastructure. It is crucial to have clearly defined standards to ensure reliable, efficient and secure communication throughout the system \cite{gungor2011smart}. The different network types in the communications layer mentioned above have all different requirements when it comes to data rate and coverage distance, and the chosen communication technology must support these specific requirements, which are summarized in Figure \ref{fig:sg_range} and Table \ref{tab:network_types_requirements}. The networks utilize different technologies for communication, both wireless and wired.
The premise network (HAN, NAN or IAN) is closest to the end-user, and enables information and communication flow between home appliances or for example heating, ventilation and air conditioning (HVAC) systems within the end-user premise. Multiple HANs connects to a NAN. The NAN collects information, and enables communication to the WAN. WAN handles communication of metering information from the end-user to the utilities companies \cite{fan2012smart}.\\

\begin{figure}[H]
    \centering
    \includegraphics[width=15cm]{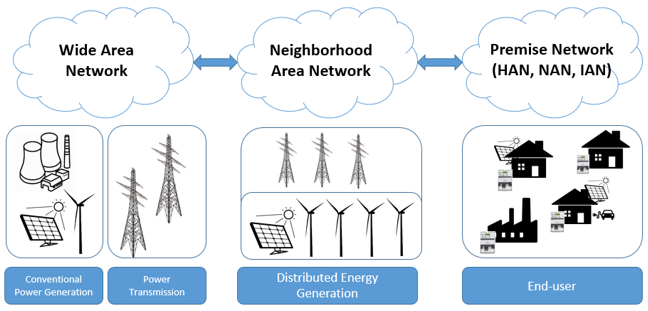}
    \caption{Networks in SG}
    \label{fig:sg_networks}
\end{figure}

\begin{table}[H]
\caption{Overview Network types and requirements}
\label{tab:network_types_requirements}
\centering
\begin{tabular}{lllcl}
\toprule
\textbf{Network type} & \textbf{Coverage} & \textbf{\begin{tabular}[c]{@{}l@{}}Data rate\\ requirements\end{tabular}} & \textbf{Data rate} & \textbf{\begin{tabular}[c]{@{}l@{}}Technology\\ alternatives\end{tabular}}\\
\midrule
WAN & 10 - 100 km & \begin{tabular}[c]{@{}l@{}}High data rate.\\ Devices such as \\ routers and\\ switches\end{tabular} & \multicolumn{1}{c}{\begin{tabular}[c]{@{}c@{}}10 Mbps\\ - 1 Gbps\end{tabular}} & \begin{tabular}[c]{@{}l@{}}Wireless: WiMAX, 3G,4G,5G.\\ Wired: Ethernet, Fiber Optic\end{tabular}                        \\
NAN/FAN & 10 m - 10 km & \begin{tabular}[c]{@{}l@{}}Highly dependent\\ on node density\\ and topology\end{tabular}& \begin{tabular}[c]{@{}l@{}}100 kbps\\ - 10 Mbps\end{tabular}& \begin{tabular}[c]{@{}l@{}}Wireless: ZigBee, WiFi,\\ WiMAX, Cellular.\\ Wired: Power Line \\ Communication\end{tabular} \\
\begin{tabular}[c]{@{}l@{}} HAN/BAN/IAN \end{tabular}& 1 - 100 m& \begin{tabular}[c]{@{}l@{}}Dependent on\\ application.\\ Generally low \\ data rate \\ required.
\end{tabular} & 10 - 100 kbps & \begin{tabular}[c]{@{}l@{}}Wireless: ZigBee, Z-wave, \\ WiFi.\\ Wired: Ethernet, HomePlug, \\M-Bus. \end{tabular} \\ \bottomrule
\end{tabular}
\end{table}

\noindent\textbf{Wide area network}\\
A WAN forms the backbone of the communication network in the power grid. It connects smaller distributed networks such as transmission substations, control systems and protection equipment e.g. supervisory control and data acquisition, remote terminal unit (RTU) and phasor measurement unit (PMU) to the utility companies’ control centers \cite{kuzlu2014communication,wang2011survey}. Other terms used for WAN is backbone network or metropolitan area network \cite{kuzlu2014communication}. WAN applications require a higher number of data points at high data rates (10 Mbps – 1 Gbps), and long-distance coverage (10 – 100 km). Real-time measurements are taken throughout the power grid by measurement and control devices and sent to control centers. In reverse, instructions and commands are sent from control centers to the devices \cite{wang2011survey}. This communication requires both a high degree of distance coverage and speed to maintain stability. Suitable communication technologies for this application are PLC, fiber optic communication, cellular or WiMAX. Satellite communication can be used as backup communication or in remote locations \cite{kuzlu2014communication,saputro2012survey}.\\
\newline
\noindent\textbf{Neighborhood area network/Field area network:} \\
The NAN and FAN are networks within the distribution domain, both enables the flow of information between WAN and a premise area network (e.g., home area network (HAN), building area network (BAN), and industrial area network (IAN)). The NAN connect premises networks within a neighborhood via smart meters at the end-user. The NAN enable services such as monitoring and controlling electricity delivery to each end-user, demand response and distribution automation.
The area NAN/FAN covers can in some cases be large, one of the features of NAN/FAN is communication between intelligent electronic devices (IEDs). The data in a NAN/FAN is transmitted from a large number of sources to a data concentrator or substation. This requires a high data rate and large coverage distance. The existing grid infrastructure in the NAN/FAN covered areas are in most cases not possible to make extensive alterations to the infrastructure.
Because of the varying nature of the physical environment of which the NAN/FAN operate, coverage requirements etc. different technologies for communication are used. When the coverage requirements are lower, standards from NAN can be applied, if longer coverage is required, other technologies will be more suitable. The communication technologies used has therefore to be adapted to each specific situation. Both wired and wireless technologies are used in NAN/FAN, and the different communication technologies should be complementary. As distributed energy generation are deployed, these are connected to the NAN/FAN. Communication technologies such as ZigBee, WiFi, Ethernet or PLC are widely used in these networks \cite{kuzlu2014communication,saputro2012survey,meng2014smart}.\\
\newline
\noindent\textbf{Premise area network}\\
The premise area network divides into three sections depending on the environment. These are wired or wireless networks within the end-user’s premise.
The purpose of the HAN is to provide communication between for example the smart meter and home automation, appliances, Home Energy Management Systems, solar panels, or electric vehicles.
BAN and IAN are commercial and industrial focused and communicate typically with building automation systems such as heating and ventilation or energy management systems. These applications do not require large coverage, high speed, or high data rate, and can be managed with low power, low-cost technologies such as PLC, WiFi or ZigBee \cite{saputro2012survey}.
The required bandwidth in HANs vary from 10 to 100 kbps for each device, depending on function. The premise networks should be expandable to allow for the number of connected devices to increase \cite{mohassel2014survey}.
Other applications for the smart metering devices within the premise area are delivering information such as power and real-time price information to the end-user through HEMS. The end-user can then make decisions whether to use appliances during high price periods or wait for lower price. This can in turn help with peak demand reduction and load shifting \cite{aravind2015smart}.

\subsection{Smart Grid Communication Technologies}
Communication technologies utilized in smart grid can as mentioned be wired or wireless. Most power systems use a combination of different wired and wireless technologies, depending on the infrastructure.
Several factor that has to be taken into account when deciding on communication technology used in SG and smart metering. Wireless communication alternatives have some advantages over wired communication, such as low cost and connectivity in inaccessible areas. A number of factors have to be considered for each different case to decide on communication technology. Requirement include aspects such as geographical topography, technical and operational requirements and cost \cite{parikh2010opportunities}.
Wireless communication is less costly to implement in a complex infrastructure and ease of installation in some areas. Wired connection will not necessarily struggle with interference issues as wireless solutions may do. Both types of communication are necessary in smart grid environment. The technology that fits one environment may not be suitable in a different environment.  Following is a summary of some of the wired and wireless communication technologies used for smart grids, together with advantages and limitations.

\subsubsection{Wired Communication}
\noindent\textbf{Power line communication (PLC)}\\
Power line communication utilizes the power transmission lines to transmit data. High frequency signals from a few kHz to tens of MHz are transferred over the power line \cite{lee2011amr}. Initial cost of PLC is lower since it uses already existing power line infrastructure. The technology is mature, and has already been in use for decades for commercial broadband and is highly reliable. PLC provide high throughput and low latency which makes it a suitable for SG communication in densely populated areas \cite{emmanuel2016communication}.
Power line communications divides into narrowband and broadband PLC. Narrowband PLC (NB-PLC) is operating at 300-500 kHz with a data rate up to 10-500 Kbps and a range up to 3 km. This is further divided into Low Data Rate Narrowband PLC and High Data Rate Narrowband PLC. Low data Rate Narrowband PLC is single carrier based, with a data rate up to 10 kbps. High Data Rate Narrowband PLC is multi carrier based with a data rate up to 1 Mbps. Broadband PLC (BB-PLC) is operating between 1.8 and 250 MHz with a data rate up to 300 Mbps. Power Line Communication can be used in nearly all parts of a smart grid environment, from home appliances in low voltage to grid automation in high voltage \cite{ai2016capacity}. Channel noise from power electronics is a major concern with this form of communication \cite{mathur2020performance, pinomaa2013noise, ai2017game}. Data distortion around transformers, and the need to bypass these using other communication techniques is another disadvantage with PLC \cite{mohassel2014survey}. Extensive field measurements show that the characteristics of PLC channel differ significantly from one environment to another, which leads to varying performance \cite{ai2017performance}. The omnipresence of the power cables also makes the combination of PLC technique with other communication technologies (e.g., RF, visible light communication (VLC), etc.) an attractive approach to extend the communication coverage, thus enabling a variety of applications such as smart home, Internet of Things, etc \cite{ai2020performancecascaded, song2015indoor, mathur2018performance}. HomePlug is type of power line communication specifically developed in-home applications and appliances. HomePlug Green PHY (HPGP) uses PLC technology, and is developed and marketed towards HAN applications. It has a data rate up to 10Mbps, and operates in the 2 MHz - 30 MHz spectrum \cite{hazen2008technology,pinomaa2015homeplug}\\
\newline
\noindent\textbf{Fiber optical communication}\\
Fiber optical communication is well suited for control and monitoring, and backbone communication in WANs, although more expensive than other alternatives it has the advantages of long range, high bandwidth and high data rate, and and not being susceptible to electromagnetic disturbances. Limitations of fiber optic communication is the number of access points. Fiber optics are used to connect substations to the utility companies control centers \cite{bian2014analysis,faheem2018smart}.\\
\newline
\noindent\textbf{Ethernet}\\
Suited for communication in WAN between substations and control centers. Advantages with this form of communications is its high availability and high reliability. Ethernet is also used in HAN for the communication between smart meters and home central.

\subsubsection{Wireless Communication}
\noindent\textbf{Cellular communication}\\
Cellular communication can be used where continuous communication is not required. It is used for communication in smart meters in rural areas. Advantages with this technology is that is already existing, it has widespread coverage, low cost and high security. One disadvantage with cellular communication is the fact that the network is shared with many other users, this can in some cases result in network congestion. Universal mobile telecommunications system (UMTS), long-term evolution (LTE), LTE-machine type communication (LTE-M) and narrowband internet-of-things (NB-IoT) are technologies used for communication in smart grids. The last two specifically developed for IoT applications. LTE-M and NB-IoT are both low power wide area networks. LTE-M offers higher data rate, but require more bandwidth \cite{karagiannis2014performance, surgiewicz2014lte, cheng2011feasibility}. \\
The fifth generation mobile communication network (5G) utilizes wide frequency range including mm wave spectra and operate at higher frequencies than LTE/4G system, allowing for higher speeds and lower latency, and ability to connect a high number of devices. This makes it suitable for SG infrastructure \cite{dragivcevic2019future,de2019security}. A comprehensive review on the use of 5G for SG with the future roadmaps and challenges is provided in \cite{dragivcevic2019future}. The security for smart grid in future generation (5G and beyond) mobile networks is discussed in \cite{de2019security}.
\\
\newline
\noindent\textbf{WiMAX (IEEE 802.16)}\\
Worldwide inter-operability for microwave access (WiMAX) is a short range wireless communication technology based on the IEEE 802.16 standards with a data rate up to 70 Mbps and a range of 50 km. WiMAX operate on two frequency bands, 11-66 GHz for line-of-sight, and 2-11 GHz for non-line of sight communication \cite{mao2012wimax}.
The physical and media access control (MAC) layer of WiMAX are defined by IEEE 802.16. The physical layer uses orthogonal frequency-division multiple access (OFDMA) and multiple-input multiple-output (MIMO) antenna system providing increased non-line of sight capabilities. The  MAC layer applies data encryption standard (DES) and advanced encryption standard (AES) encryption to ensure secure and reliable communication. The MAC layer also utilizes power saving techniques such as sleep and idle \cite{usman2013evolution}.
WiMAX is scalable and can be set up as networks on local or regional level. WiMAX is well suited for sensors and meters provided sufficient numbers of nodes in the area. One limitation with WiMAX is that coverage becomes highly limited due to signal losses (e.g., rain attenuation, blockage, etc.) \cite{rengaraju2012communication}.
\\
\newline
\noindent\textbf{ZigBee (IEEE 802.15.4)}\\
ZigBee is an open wireless mesh network standard based on the IEEE 802.15.4 standard. It is a short range, low data rate, and energy efficient technology. ZigBee operate on four different frequency bands, 868cMHz (20 kbps per channel), 915 MHz (40 kbps per channel), and 2.4 GHz (250 kbps per channel) \cite{parvez2014frequency,mahmood2015review}. ZigBee has mesh capabilities and a coverage range from 10 to 100 meters \cite{gungor2011smart}. Mesh networks are decentralized, where each node are self-manageable, and can re-route, and connect with new nodes when needed.  This makes ZigBee well suited for use in HAN applications such as remote monitoring, home automation, consumer electronics and smart meter readings \cite{zafar2018prosumer,burunkaya2017smart}.
A ZigBee mesh network is constructed of three different types of nodes: coordinator, router and end-device. ZigBee uses AES-128 access control to manage a high level of security. Because of the low transmission power level, this technology is vulnerable to multipath distortion, noise and interference \cite{mahmood2015review}. ZigBee operating on 2.4 GHz band is also affected by distortion from technologies such as WiFi, USB, Bluetooth and microwave ovens as these operate on the same unlicensed frequency band \cite{ai2015radio, cheffena2016industrial, ai2015power}.
\\
\newline
\noindent\textbf{Z-Wave (IEEE 802.15.4)}\\
Z-Wave is a proprietary communications standard intended for remote control of applications in residential and commercial areas. In Europe, it operate on 868 MHz with a data rate of 9.6 kbps, and on 2.4 GHz with a data rate up to 200 kbps. Range wise Z-Wave typically have around 30 meters indoor range, and up to 100 meters outdoors. Z-Wave is short range, low data rate and low cost alternative. Z-Wave can also be organized as mesh network, increasing the range \cite{mahmood2015review}. Similar to ZigBee, Z-Wave also uses AES128 encryption standard to maintain a high level of security in the network \cite{bekara2014security}.
\\
\newline
\noindent\textbf{WiFi (IEEE 802.11)}\\
WiFi technology, based on the IEEE 802.11 family of standards, is a wireless networking technique that is being widely used for Internet access. It can also be a good choice in the context of smart grid, which enables consumers to monitor the improve their energy use \cite{samuel2016review}. WiFi solution is already being utilized in a number of devices that contribute to the so-called smart home. For instance, WiFi is used in thermostats, appliances, and new smart energy home devices that will connect them all together to help consumers manage their own energy consumption \cite{granelli2014usage, li2011applications, samuel2016review}.
\\
\newline
\noindent\textbf{Satellite communication}\\
Satellite communication can play an important role in SG communication in rural areas without cellular coverage, or as a backup solution for other communication technologies \cite{meloni2017role}. Examples of areas of use for satellite communication are control and monitoring of remotely located substations \cite{de2015satellite}.
\\
\newline
\noindent\textbf{Free-space optical (FSO) communications}\\
The demand for higher data rates requires broader bandwidth for communication system. Among different potential technologies, FSO communication is one of the most promising technologies addressing the problem of large bandwidth and data rate requirements, as well as the “last mile bottleneck”. FSO system functions by transmitting modulated laser light through the air between the transmitter and receiver. More specifically, the signal is transmitted using a lens or parabolic mirror by narrowing the light and projecting it towards the receiver. The emitted light is then picked up at the receiver with a lens or mirror. Subsequently, the received light is focused on an optical detector and converted to electrical signals for further information extraction \cite{fredrik2020fso}. Besides the advantages of large data rate with unlicensed spectrum, the FSO communication is also considered to be a more secure technique than the RF communication \cite{ai2019physical, gyan2020secrecy, ai2020secrecy, ai2020comprehensive}. Thanks to the various advantages of FSO communications, FSO link can be part of the backhaul communication network for rural or remote substations monitoring applications. In \cite{sivakumar2020novel}, the FSO system based on microring resonator (MRR) with the ability to deliver up to gigabit (line-of-sight) transmission per second is proposed for the two smart grid applications (AMI and DR).  The experimental results demonstrate up to 10 times bandwidth improvement over the radius as large as 600 meters and maintain receive power higher than the minimum threshold (- 20 dBm) at the controller/users, so the overall system is still able to detect the FSO signal and extract the original data without detection. The feasibility of FSO communications technology from the atmospheric context of Bangladesh has been analyzed for smart village energy autonomous systems in \cite{haider2018approach}.

\begin{landscape}
\begin{table}[H]
\caption{Overview of wired communication technologies in SG.}
\label{tab:wired_comm_tech}
\centering
\begin{tabular}{lllllll}
\toprule
\multicolumn{7}{l}{Wired communication technologies}\\
\midrule
Technology & Data rate & Coverage & Application & Advantages & Disadvantages & Network type\\
\midrule
Ethernet       & \begin{tabular}[c]{@{}l@{}}Up to\\ 100 Gbps\end{tabular}     & \begin{tabular}[c]{@{}l@{}}Up to \\ 100 m\end{tabular} & \begin{tabular}[c]{@{}l@{}}In-home communication,\\ SCADA, backbone\\ commnunication\end{tabular} & \begin{tabular}[c]{@{}l@{}}Good on short \\ distances\end{tabular} & \begin{tabular}[c]{@{}l@{}}Coverage\\ limitations\end{tabular} & \begin{tabular}[c]{@{}l@{}}Premise network,\\ NAN/FAN, WAN\end{tabular} \\

Broadband PLC  & \begin{tabular}[c]{@{}l@{}}Up to \\ 300 Mbps\end{tabular}   & \begin{tabular}[c]{@{}l@{}}Up to\\ 1500 m\end{tabular} & \begin{tabular}[c]{@{}l@{}}SCADA, backbone\\ communication in \\ power generation \\ domain\end{tabular} & \begin{tabular}[c]{@{}l@{}}Existing\\ infrastructure,\\ standardized,\\ high reliability\end{tabular} & \begin{tabular}[c]{@{}l@{}}Noisy channel\\ environment,\\ Disturbance\end{tabular}                              & NAN/FAN, WAN \\

Narrowband PLC & \begin{tabular}[c]{@{}l@{}}10-500\\ Kbps\end{tabular}       & \begin{tabular}[c]{@{}l@{}}Up to\\ 3 km\end{tabular}   & \begin{tabular}[c]{@{}l@{}}SCADA, backbone\\ communication in power\\ generation domain\end{tabular} & \begin{tabular}[c]{@{}l@{}}Existing\\ infrastructure,\\ standardized,\\ high reliability\end{tabular} & \begin{tabular}[c]{@{}l@{}}Noisy channel\\ environment,\\ Disturbance\end{tabular} & NAN/FAN, WAN \\

HomePlug       & \begin{tabular}[c]{@{}l@{}}4, 5, 10\\ Mbps\end{tabular}     & \begin{tabular}[c]{@{}l@{}}Up to\\ 200 m\end{tabular}  & \begin{tabular}[c]{@{}l@{}}In-home communication,\\ Smart appliances\end{tabular} & \begin{tabular}[c]{@{}l@{}}Low cost, \\ low energy\end{tabular} & \begin{tabular}[c]{@{}l@{}} Coverage \\ limitations, \\Disturbance \end{tabular} & Premise network \\

Fiber optic    & \begin{tabular}[c]{@{}l@{}}Up to\\100 Gbps\end{tabular} & \begin{tabular}[c]{@{}l@{}}Up to\\ 100 km\end{tabular} & \begin{tabular}[c]{@{}l@{}}SCADA, backbone \\ communication in power\\ generation domain\end{tabular} & \begin{tabular}[c]{@{}l@{}}High bandwidth,\\ high data rate.\\ not susceptible to\\ electromagnetic\\ interference\\\end{tabular} & Costly & WAN \\
\midrule
\end{tabular}
\end{table}
\end{landscape}

\begin{landscape}
\begin{table}[H]
\caption{Overview of wireless communication technologies in SG.}
\label{tab:wireless_comm_tech}
\centering
\begin{tabular}{lllllll}
\toprule
\multicolumn{7}{l}{Wireless communication technologies} \\
\midrule
Technology& Data rate& Coverage& Application& Advantages& Disadvantages& Network type \\
\midrule
WiMAX& 75 Mbps & \begin{tabular}[c]{@{}l@{}}Up to\\ 50 km\end{tabular}  & \begin{tabular}[c]{@{}l@{}}In-home communication\\ Smart meter reading\end{tabular} & \begin{tabular}[c]{@{}l@{}}Low cost,\\ low energy\end{tabular} & \begin{tabular}[c]{@{}l@{}}Not widespread,\\ coverage highly\\ reduced if loss in \\ line of sight\end{tabular} & NAN/FAN, WAN \\
ZigBee & 20-250 kbps & \begin{tabular}[c]{@{}l@{}}Up to\\ 100 m\end{tabular} & \begin{tabular}[c]{@{}l@{}}In-home communication,\\ energy monitoring,\\ smart appliances,\\ home automation\end{tabular} & \begin{tabular}[c]{@{}l@{}}Mesh capability,\\ simplicity, mobility,\\ low energy, low cost.\end{tabular} & \begin{tabular}[c]{@{}l@{}}Low data rate,\\ short range,\\ interference\end{tabular} & \begin{tabular}[c]{@{}l@{}}Premise network,\\ NAN/FAN\end{tabular} \\
Z-Wave & 9-40 kbps & \begin{tabular}[c]{@{}l@{}}Up to\\ 30 m\end{tabular} & Wireless mesh network & \begin{tabular}[c]{@{}l@{}}Mesh capability,\\ simplicity, mobility,\\ low energy, low cost.\end{tabular} & \begin{tabular}[c]{@{}l@{}}Low data rate,\\ short range,\\ interference\end{tabular}  & Premise network \\
WiFi & \begin{tabular}[c]{@{}l@{}}2 Mbps -\\ 1.7 Gbps\end{tabular} & \begin{tabular}[c]{@{}l@{}}Up to\\ 100 m\end{tabular}  & \begin{tabular}[c]{@{}l@{}}In-come communication,\\ smart appliances,\\ home automation,\\ SCADA\end{tabular} & \begin{tabular}[c]{@{}l@{}}Good on short\\ distances.\end{tabular} & Security & \begin{tabular}[c]{@{}l@{}}Premise network,\\ NAN/FAN\end{tabular} \\
3G & \begin{tabular}[c]{@{}l@{}}Up to\\ 42 Mbps\end{tabular} & 70 km & \begin{tabular}[c]{@{}l@{}}SCADA,\\ Smart meter reading\end{tabular} & \begin{tabular}[c]{@{}l@{}}Already existing \\ network, high\\ security, low cost, \\ large coverage\end{tabular} & \begin{tabular}[c]{@{}l@{}}Network shared\\ with consumers\\ may result in\\ congestion.\end{tabular} & NAN/FAN, WAN \\
4G/LTE & \begin{tabular}[c]{@{}l@{}}Up to\\ 979 Mbps\end{tabular} & \begin{tabular}[c]{@{}l@{}}Up to\\ 16 km\end{tabular} & \begin{tabular}[c]{@{}l@{}}SCADA,\\ Smart meter reading\end{tabular} & \begin{tabular}[c]{@{}l@{}}Already existing\\ network, high\\ security, low cost,\\ large coverage\end{tabular} & \begin{tabular}[c]{@{}l@{}}Network shared\\ with consumers\\ may result in\\ congestion.\end{tabular} & NAN/FAN, WAN \\
LTE-M & 7 Mbps & 11 km & Smart meter reading & \begin{tabular}[c]{@{}l@{}}Low cost, low\\ energy, scalability,\\ coverage\end{tabular} & Lower data rate & NAN/FAN \\
NB-IoT & 159 kbps & & Smart meter reading & \begin{tabular}[c]{@{}l@{}}Low cost, low\\ energy, scalability,\\ coverage\end{tabular} & Lower data rate & NAN/FAN \\
5G & \begin{tabular}[c]{@{}l@{}}Up to\\ 20 Gbps\end{tabular} &  & \begin{tabular}[c]{@{}l@{}}SCADA\\Smart meter reading \\ \end{tabular} & \begin{tabular}[c]{@{}l@{}}High data rate, \\ scalability\end{tabular} & Lower data rate & NAN/FAN, WAN \\
Satellite & 50 Mbps & & \begin{tabular}[c]{@{}l@{}}Backup, remote location\\ communication\end{tabular} & \begin{tabular}[c]{@{}l@{}}Good when no\\ other alternative is\\ viable\end{tabular} & High cost & WAN \\
\bottomrule
\end{tabular}
\end{table}
\end{landscape}

\section{Challenges of smart grid communication}

In this section we will discuss future trends of smart grid communications and applications, and a comprehensive review of these challenges.

\subsection{Robust Transmission}
Robust transmission of information with high QoS is one of the most prioritized requirements for smart grid communications. It will greatly improve the system robustness and reliability by harnessing the modern and secure communication protocols, the communication technologies, faster and more robust control devices, embedded intelligent devices for the entire grid from substation and
feeder to customer resources \cite{moslehi2010smart}. As the use of communication systems in other scenarios, there are many challenges to achieve robust transmission because of limited bandwidth, limited power, adverse transmission environment (interference, high path loss, etc) \cite{ai2015comparative, guzelgoz2011review, ai2017path, al2020rain, ai2017multi}. As discussed in the previous sections, both wireless and wired communication technique consists important parts of the smart grid communication with its own advantages and disadvantages. In many cases, a hybrid communication technology mixed with wired and wireless solutions can be used in order to provide higher level of system reliability, robustness and availability\cite{zhang2018hybrid}.

\subsection{Security}

Cyber security is considered to be one of the biggest challenges to smart grid deployment as the grid becomes more and more interconnected, and every aspect of the SG must be secure \cite{yan2012survey}. Security measures must cover issues involving communication and automation that affects operation of the power system and the utilities managing them. It must address deliberate attacks as well as inadvertent accidents such as user error and equipment failure \cite{nist_roadmap_3,yan2012survey}.

Smart grids are vulnerable to cyber-attacks due to the integration of communication paths throughout the grid infrastructure. Smart grids are still evolving, and considering security in a new smart grid environment is important, but challenging. Undetected cyber-attacks can lead to critical damage affecting thousands or millions of customers and life threatening infrastructure \cite{saxena2018impact,yang2011communication}.
Securing the data is vital for both end-user and power companies to ensure trust. As more functions and capabilities are implemented to the smart grid importance of secure and safe communication increase. From distributed energy generation, energy storage, electric vehicles to power station and power grid control systems. Also something possibly as trivial as securing that the reading from the end-user’s smart meters are sending correct billing information, or that the utilities companies receive the correct information is essential \cite{sgouras2014cyber}. As for any other communication systems, security enhancement for smart grid communication can be achieved at different layer of the protocol by utilizing the techniques from the conventional upper layer cryptography \cite{iyer2011cyber, he2016lightweight, gope2018privacy, sharma2016survey, nicanfar2014password} to the physical layer security \cite{lee2012physical, ai2019secrecy, islam2019physical, ai2019secrecyel}.
Different communication technologies, wired and wireless, interconnects and are required to operate the grid securely. Different authorities are responsible securing different data and security aspects in smart grid/smart metering. For the Norwegian case:
\begin{itemize}
    \item Norwegian metrology service: measurement accuracy
    \item Norwegian directorate for civil protection (DSB): electrical safety
    \item Norwegian communications authority: communication
    \item Norwegian water resources and energy directorate: application, function and safety of smart meters.
\end{itemize}
Personal data act, Section 13 sets requirements for satisfactory information security \cite{personaldataact}. Based on this, the Norwegian Electrotechnical Committee (NEK) emphasizes on the following three aspects in relation to security in smart grids: confidentiality, integrity and availability, as well as the following four elements \cite{nek_amsplushan_gjoretilgjengelig}.
\begin{itemize}
    \item Protection against unauthorized access to measurement data on the meter.
    \item Protection against unauthorized retrieval of measurement data.
    \item Protection against tampering or alteration of measurement data.
    \item Ensuring that measurement data is available when needed.
\end{itemize}

Vulnerabilities and threats may also be categorized as consumer threat, naturally occurring threat, individual and organizational threat, impacts on consumer, impacts on availability, financial impacts and likelihood of attack \cite{tan2017survey}. NEK recommends that communication in HAN use synchronous encryption AES-128, since the data has fixed length. The end-user have to request the utility company to open up for HAN, and to receive encryption key \cite{nek_2018_han_personvern}.
In Norway, the Norwegian Data Protection Authority (DPA) has identified several aspects relating to smart grid and smart metering privacy. Since the smart meters can be linked to address, and home-owner, behavioral information can be traced back to individual person \cite{datatilsynet2018automatisk}.
Earlier SCADA systems were isolated on a separate computer network, but the development towards connecting all devices to the Internet, are making the system vulnerable to cyber-attacks \cite{kumar2018developing, sun2018cyber}. Attacks have been carried out on SCADA networks in the past, some with significant impact to infrastructure and power delivery \cite{liang20162015}.
Attacks on smart grids can occur on all levels, from generation and distribution to home networks, it can be protocol-based attacks, routing attacks, intrusion, malware and denial-of-service attacks (DoS). Successful attacks can lead to grid instability, or worst case failure and blackouts \cite{sgouras2014cyber,tan2017survey,kim2012survey}.
A reliable SG depends on avoiding attacks, or detecting and establishing mitigation measures. Protection used within SG for message authentication, integrity and encryption. Security must also address loss of communication, unauthorized access to network and devices (eavesdropping), network attacks denial of service (DoS), distributed denial of service (DDoS), man-in-the-middle (MITM), and jamming of radio signals \cite{de2019security}.

There have been several attacks on power companies the last years, where some have led to system failure and blackout.
In 2006 a nuclear power plant in Alabama, USA failed due to overload on the control system network. Investigations later identified the source to be manipulated smart meter power readings \cite{kim2012survey}.
In 2013-2014 an attack affected more than 1000 energy companies in 84 countries including Germany, France, Italy, Spain, Poland and the US \cite{kshetri2017hacking}.
In December 2015, Ukraine experienced a cyber attack on three regional power distribution companies, leaving people in the dark for over six hours. Over two months after the attack, control centers were not fully operational. The attack was distributed via spear-phishing email, targeting IT staff and systems administrators in companies responsible for power distribution. By opening an attachment in an email, malicious firmware were uploaded SCADA-network. The intruders gained access to substation control centers via virtual private networks (VPNs) and was able to send commands to disable uninterruptible power supply (UPS) systems, and open breakers in substations. The blackout affected around 225,000 customers, and manual operations were required to turn the power back on \cite{saxena2018impact,liang20162015}.
In 2016 Ukrainian power distribution was once again attacked, parts of the city of Kyiv lost power for an hour. The malware enabled control of circuit breakers to the attackers.
In 2020, the European Network of Transmission System Operations for Electricity experienced an attack on its office network. The attack did however not infect any of the systems responsible for controlling the power grid \cite{fingrid2020,kshetri2017hacking}. \\

\noindent\textbf{Denial of service attack}\\
It has been claimed that DoS attacks are one of the greatest concerns for service providers. SG consists of a number of measurement devices such as smart meters, smart appliances, data aggregators, PMUs, IEDs, RTUs, PLCs, etc. Attacks to SGs can result in loss of data availability, loss of communication control, compromised data integrity, and loss of power \cite{huseinovic2020survey}. \\

\noindent\textbf{Use of encryption}\\
The security of the power grid is depends on authentication, authorization. Encryption of communication flowing between devices in the grid and data centers is crucial to reduce attackers ability to gain access to data or achieve system control. Depending on communication technology, different solutions are preferred, such as advanced encryption standard (AES) and triple data encryption (TDES) \cite{yan2012survey}. Encryption ensures identification and authorization. \\

\noindent\textbf{Authentication and authorization} \\
Authentication is the process of verifying the identity of a user, application or device.
Authorization in the process of verifying whether the user, application or device has permission or the rights to access the system, or perform an operation. Authentication, authorization and access control is necessary in SG due to the vast amounts of connected devices. Different users with different roles and level have access to control systems, sensors communication networks in the SG. Entities in the SG must be bidirectionally authenticated.
Common types of authentication schemes in SG are: device-to-device, device-to-network and user-to-network \cite{saxena2015state}.\\

\subsection{Privacy}
Communication in smart grids are often linked to information related to individual customers and their lives. This is why securing authentication, authorization, and confidentiality is so important in a smart grid environment. It is of greatest importance not to disclose private data customer to anyone other than consented entities. Private data include consumer identification, address, energy usage information \cite{tsado2015resilient}.
Smart meters are expected to provide high accuracy reading of power consumption at defined time intervals to the utilities companies. This data is used for billing purposes and grid management. However, measurement data from smart meter may be used for other purposes. Usage pattern analysis can be useful for power saving, but involves a significant risk. The data holds a great amount of information about individual consumers \cite{fan2012smart,quinn2009privacy}. Non-intrusive appliance load monitoring (NALM) technologies uses extracts detailed information on appliance use based on energy measurements \cite{hart1992nonintrusive}. By analyzing data and usage patterns, it may be possible to predict when people are at home or away from home, or what appliances are in use. This information is could be of interest for the police, tax authorities, insurance companies etc. \cite{goel2015smart,prudenzi2002neuron}. NIST have acknowledged that the major benefit of smart grids is the ability to receive richer data from smart meters and devices, is also the biggest weakness from a privacy standpoint.\\

%%%%%%%%%%%%%%%%%%%%%%%%%%%%%%%%%%%%%%%%%%%
\section{Conclusion}
In this paper, an overview of smart grid infrastructure, communications technologies, and its requirements, and applications in premises network, neighborhood area network and wide area network were presented. Cyber security challenges are briefly presented.
We are currently in the brief beginning of what will be a major change in how electric power grids and power generation are organized and managed. The changes are likely to be significant, and new possibilities emerge as new technologies are further developed.
The amount of data and information exchange are increasing rapidly as new technologies are implemented to the grid. Security concerns must be addressed to ensure a reliable power supply.

%%%%%%%%%%%%%%%%%%%%%%%%%%%%%%%%%%%%%%%%%%
%\section{Patents}
%This section is not mandatory, but may be added if there are patents resulting from the work reported in this manuscript.

%%%%%%%%%%%%%%%%%%%%%%%%%%%%%%%%%%%%%%%%%%
\vspace{6pt}

\reftitle{References}

%\begin{thebibliography}{999}
% Reference 1
%\bibitem[Author1(year)]{ref-journal}
%Author1, T. The title of the cited article. {\em Journal Abbreviation} {\bf 2008}, {\em 10}, 142--149.
% Reference 2
%\bibitem[Author2(year)]{ref-book}
%Author2, L. The title of the cited contribution. In {\em The Book Title}; Editor1, F., Editor2, A., Eds.; Publishing House: City, Country, 2007; pp. 32--58.
%\end{thebibliography}

% The following MDPI journals use author-date citation: Arts, Econometrics, Economies, Genealogy, Humanities, IJFS, JRFM, Laws, Religions, Risks, Social Sciences. For those journals, please follow the formatting guidelines on http://www.mdpi.com/authors/references
% To cite two works by the same author: \citeauthor{ref-journal-1a} (\citeyear{ref-journal-1a}, \citeyear{ref-journal-1b}). This produces: Whittaker (1967, 1975)
% To cite two works by the same author with specific pages: \citeauthor{ref-journal-3a} (\citeyear{ref-journal-3a}, p. 328; \citeyear{ref-journal-3b}, p.475). This produces: Wong (1999, p. 328; 2000, p. 475)

%=====================================
% References, variant B: external bibliography
%=====================================
\externalbibliography{yes}
\bibliography{bibliography}

%%%%%%%%%%%%%%%%%%%%%%%%%%%%%%%%%%%%%%%%%%
%% optional
%\sampleavailability{Samples of the compounds ...... are available from the authors.}

%% for journal Sci
%\reviewreports{\\
%Reviewer 1 comments and authors’ response\\
%Reviewer 2 comments and authors’ response\\
%Reviewer 3 comments and authors’ response
%}

%%%%%%%%%%%%%%%%%%%%%%%%%%%%%%%%%%%%%%%%%%
\end{document}